\documentclass[12pt]{article}
\usepackage{astrobib,epsf,aaspp4,psfig,lscape}
\def\lap{\lower.5ex\hbox{$\; \buildrel < \over \sim \;$}}
\def\gap{\lower.5ex\hbox{$\; \buildrel > \over \sim \;$}}
\begin{document}
\title{Identification and Analysis of Young Star Cluster Candidates in M31\altaffilmark{1}}
\author{ Benjamin F. Williams}
\affil{University of Washington}
\affil{Astronomy Dept. Box 351580, Seattle, WA  98195-1580}
\affil{ben@astro.washington.edu}
\author{Paul W. Hodge}
\affil{University of Washington}
\affil{Astronomy Dept. Box 351580, Seattle, WA  98195-1580}
\affil{hodge@astro.washington.edu}

\altaffiltext{1}{Based on observations with the NASA/ESA Hubble Space
Telescope obtained at the Space Telescope Science Institute, which is
operated by the Association of Universities for Research in Astronomy,
Inc., under NASA contract NAS5-26555.}

\begin{abstract}

We present a method for finding clusters of young stars in M31 using
broadband WFPC2 data from the HST data archive.  Applying our
identification method to 13 WFPC2 fields, covering an area of $\sim$60
arcmin$^2$, has revealed 79 new candidate young star clusters in these
portions of the M31 disk.  Most of these clusters are small
($\lesssim$5 pc) young ($\sim$10-200 Myr) star groups located within
large OB associations.  We have estimated the reddening values and the
ages of each candidate individually by fitting isochrones to the
stellar photometry.  We provide a catalog of the candidates including
rough approximations of their reddenings and ages.  We also look for
patterns of cluster formation with galactocentric distance, but our
rough estimates are not precise enough to reveal any clear patterns.

\end{abstract}


\section{Introduction}

Observations of extragalactic young star clusters are essential for
understanding how star formation affects galaxy evolution.  The young
stellar population is responsible for many of the characteristics
which give spiral galaxies their current morphological classification.
Although massive young stars constitute only a very small percentage
of the stellar population of most spiral galaxies, they trace the most
recent star formation, produce and disperse most of the heavy
elements, and illuminate the spiral arms. Open clusters are the
typical birthplaces of bright massive stars.  Since these clusters
contain the youngest stars in the galaxy, their identification and
examination is crucial for learning how star formation has progressed
within the galaxy, resulting in its current appearance.  The most
detailed information about these clusters comes from studies of their
constituent stars.

The study of extragalactic OB associations has been an ongoing
struggle for 5 decades, going back to studies of the properties of the
very conspicuous associations of bright stars in the Large Magellanic
Cloud (LMC) in the 1950s (e.g. \citeNP{Buscombe}, \citeNP{Shapley}).
Decades of research have resulted in catalogs of OB associations in
nearby galaxies (e.g. \citeNP{lucke} (LMC), \citeNP{vandenbergh1964}
(M31), \citeNP{hodge1977} (NGC 6822)).  These catalogs have provided
excellent starting points for studying the young stellar populations
of other galaxies; however, these samples were not ideal for further
statistical analysis because they were obtained through subjective
analysis of low-resolution, non-uniform data.

More recently, high-resolution photometric data and advances in
computational analysis techniques have allowed more objective
identification of OB associations in galaxies (e.g. \citeN{wilson}),
creating more uniform samples on which to perform statistical
analyses.  These objective samples have resulted in significant
advances in our understanding of the properties of extragalactic OB
associations.  The similarity of their size distributions and stellar
luminosity functions across different galaxy types
\cite{Bresolin1998}, and their similar luminosity function across
galaxy types \cite{Batinelli2000} have suggested that massive star
formation occurs by similar processes in all galaxies.  At the same
time, OB associations' average population sizes appear to differ
between galaxies of different morphological type \cite{bresolin1997},
showing that environment has some effect on massive star formation.

M31 has been an excellent laboratory for the study of massive star
formation due to its proximity and its many active spiral arms.  These
arms contain hundreds of OB associations that have been studied
extensively from the ground and from space.  Our distant view of M31
has allowed ground-based studies of young clusters on a wide range of
size scales.  Due to the crowding of stars and variable reddening in
the disk, the optical ground based data was used mainly for
identification and studies of the structure of OB associations.  These
studies have shown, for example, that massive star formation appears
hierarchical \cite{battinelli1996}: the large complexes contain many
smaller clusters, possibly with related physical properties.  In the
infrared, the crowding and reddening effects are reduced.  Recent
infrared work has succeeded in finding evidence for episodic star
formation which occurred in different subregions during each spiral
wave passage \cite{kodaira1999}.  Space-based observations are
allowing more detailed analysis of these regions of recent star
formation.  \citeN{magnier1997} have used Hubble Space Telescope (HST)
photometric data to determine the reddening distribution and ages of a
handful of OB associations.

While studies of the well-known OB associations in M31 have been
instrumental to our understanding of massive star formation, there has
been very little work done on the small, young and compact star
clusters in the spiral arms of M31.  These clusters have been
difficult to identify from the ground since they require very high
resolution to resolve.  \citeN{hodge1979} discovered several hundred
open cluster candidates in M31, but these objects were larger than
typical Galactic clusters and may actually be small OB associations.
In this study, we use the resolving power of HST in order to probe the
size scales of typical open clusters in M31.  We create a cluster
finding algorithm fine-tuned to find small young clusters in the M31
disk and identify their most likely member stars.  We then use the
photometry from these stellar populations to estimate reddening and
age values for our sample of young cluster candidates.

\section{Data Acquisition and Reduction}
 
We searched the HST archive for all exposures of longer than 60
seconds which were taken through the broadband B (F439W) and V (F555W)
filters pointing within 1.5 degrees of the M31 nucleus.  Using this
narrow radius kept our data contained within the disk, avoiding
significant halo contamination.  Any fields in the bulge were later
excluded by eye.  We acquired and reduced 13 WFPC2 fields from the HST
archive, each observed through at least the B and V broad-band
filters.  U band (F336W) images were also available for most of these
fields, providing useful information about the reddening of our
cluster candidates.  Table 1 gives the RA, DEC, dates, bandpasses, and
exposure times of the data taken from the HST archive in order to put
together each of these fields. The positions of these fields on the
galaxy are shown in Figure 1.  The figure shows an H$\alpha$ mosaic of
the M31 disk \cite{winkler1995} with the positions of the 13 fields
marked with squares showing the positions of each chip in the field.
Each field is labeled with the number given in Table 1. The PC chip
position is not shown because it was excluded from our analysis.
Since the PC portions of the images often contained special regions of
the galaxy (e.g. globular clusters) and since the area covered by the
PC is small, we decided to exclude all of the PC data in order to keep
the data set as unbiased as possible.  The exposure times for these
fields were generally quite short, and therefore they are relatively
shallow.

We determined instrumental magnitudes for each of these fields using
the automated programs DAOPHOT II and ALLSTAR \cite{stetson}.  An
object was considered a real star if it was detected as a point source
at 4$\sigma$ above the noise level in 2 bandpasses with centroids
separated by less than 0.7 pixels.  Therefore, some of the objects
under consideration may be misclassified background galaxies and
foreground stars; however, in fields of these angular sizes at this
galactic latitude (-21.57$\deg$) and these depths, the average number
of foreground galactic stars per field should be less than 10.
Background galaxies will typically not be detected as they will not
appear as point sources.  Figures 2 and 3 show the typical photometric
errors, determined by ALLSTAR, as a function of B and V magnitude for
each of the 13 fields.  The errors tended to be approximately at the
10\% level near the bright magnitudes ($m_V<24$), mostly due to the
fluctuating surface brightness of the background.  The M31 disk is a
complex background with which to work and therefore limits the
accuracy of the photometry by increasing the uncertainty of the local
background level.  This uncertainty is partially due to the poisson
noise of the higher background levels, but it is also due to actual
structure in the background on spatial scales relevant to stellar
photometry.  These surface brightness fluctuations come from structure
in the stellar disk which cannot be resolved by HST.

Point Spread Function (PSF) magnitudes were checked against aperture
photometry of the most isolated stars with the highest signal to noise
in order to determine if there was an offset between the PSF
photometry and the more standard aperture photometry.  We found small
offsets of our PSF photometry from the aperture photometry on the WF3
chip in the V band and in the WF4 chip in the B band.  We applied
small corrections (+0.03 mags for stars measured in the V band on the
WF3 chip, and +0.05 mags for stars measured in the B-band on the WF4
chip) to our photometry in these cases in order to make the mean
offset between the aperture and PSF photometry zero.  In all other
cases, the offsets were zero.  We then obtained standard U, B, and V
magnitudes from our instrumental magnitudes using the methods, zero
points, and transformation coefficients given in \citeN{holtzman1995}.
We first determined instrumental magnitudes for the F336W, F439W, and
F555W filter as
$$ 
m_{filter} = -2.5 \times log(ADU/t) + X + ZP_{filter}\\ \ \ \ \ \ (1) 
$$ 
where ADU is the number of counts, t is the exposure time, X is the
small offset computed from the aperture photometry mentioned above, and
$ZP_{filter}$ is the zero point of the WFPC-2 chip for the bandpass.
Since the transformation to U, B, and V is a function of color, we
used the F336W - F439W as a first approximation of the U-B color and
the F439W - F555W color as a first approximation of the B-V color and
iteratively solved the transformation equations.

As a further test to the accuracy of our photometry, we took advantage
of the fact that two of our fields were overlapping.  The WF3 chip in
field 7 was covering the same region of space as the WF2 chip in field
9.  We used the IRAF\altaffilmark{2} tasks GEOMAP and GEOTRAN to determine a
rotational and translational conversion between the coordinates of the
stars in one frame to their coordinates in the other frame.  We were
then able to convert all of the pixel coordinates of the stars in one
frame to the coordinates of the same star in the other frame.  By
comparing the star lists, we found every star which was detected in
both frames.  Then we were able to compare independent measurements of
the same stars as determined in different locations on the chips, on
different chips and in different frames.

\altaffiltext{2}{IRAF is distributed by the National Optical Astronomy
Observatories, which are operated by the Association of Universities
for Research in Astronomy, Inc., under cooperative agreement with the
National Science Foundation.}

The easiest way to see the accuracy of our photometry is by looking at
the residuals after subtracting the magnitudes of the stars determined
on the WF3 chip from the same stars observed with the WF2 chip.  These
residuals are shown in Figure 4.  No systematic differences between
the chips are seen, and the residuals are consistent with zero in
nearly all cases.  With this reinforcement that we understood our
errors, and with the knowledge that our photometry and errors were
accurate, we could apply our young cluster finding algorithm to the
stellar photometry.

\section{Identification of Young Clusters of Stars}

In order to explore the many open clusters and young stellar
associations, we created an objective method for detecting the
clusters within the fields.  Using our stellar positions and
photometry from DAOPHOT II and ALLSTAR, the mean surface density of
bright ($m_V\ <\ 24.5$, $M_V\ \lesssim\ 0.1$) blue ($B-V < 0.45$)
stars is determined.  Then the standard deviation from the mean
density ($\sigma$) on a size scale specified by the user is
calculated.  This calculation is performed by comparing the density of
bright blue stars in regions of the specified size around every bright
blue star detected in the field with the mean density.  This
calculation of the mean stellar density and the standard deviation of
the stellar density is followed by a search for regions containing at
least 4 bright blue stars and having surface densities of bright blue
stars 3$\sigma$ above the mean.

We found the user specified size must be chosen carefully.  Large
values will often find overdensities which contain more than one
cluster, unnecessarily reducing the spatial resolution of the data.
Small values will often lead to single clusters being divided into
several stellar overdensities of the size requested.  We found the
best way to overcome these problems was to run the algorithm using two
size scales, one corresponding to the sizes of a few of the smaller
clusters visible in the images and one corresponding to the sizes a
few of the larger clusters visible in the images.  If a cluster was
found at both size scales, we had to choose which of the detections
was most appropriate to use for follow-up work.  If the cluster
appeared small and populous we would use the detection from the small
search radius in order to avoid field contamination in our star
sample.  If the cluster was large, and it had been detected as more
than one cluster with the small search radius, we would use the
detection from the large search radius in order to maximize the number
of stars in our sample and in order to avoid making multiple age and
reddening measurements for the same cluster.

Finally, in order to remove statistical anomalies from our sample, we
ran each cluster candidate through a surface brightness test.  In this
test, we measure the surface brightness around the center of each
cluster candidate.  This test was a bit more complex since, due to
completeness issues, the centers of the stellar overdensities were not
always aligned exactly with the high surface brightness regions.
Therefore we allowed overdensities with very nearby high surface
brightness regions to pass.  This step required great care for the
larger clusters since the chance of one bright star entering the
surface brightness calculation and enhancing the surface brightness
near the center of the overdensity was higher for the large clusters.
We found the number of these single bright star contaminants was
reduced when we removed candidates with substantial ($>$2 mag
arcsec$^{-2}$) increases in surface brightness away from the center.
These large jumps in surface brightness were usually bright single
point sources, which were likely foreground stars.  Any stellar
overdensity whose measured surface brightness characteristics did not
pass our objective criteria was not likely part of an underlying
cluster of unresolved stars.  These low surface brightness regions
were removed from the sample.

We ran this algorithm on our star lists for all of the WFPC2 chips for
each field in our sample.  We did one run looking for small scale
associations (radius $\sim$5 pc), and we did a second run looking for
larger associations (radius $\sim$15 pc).  We compared our results
with published catalogs of clusters and associations, and with
previously performed eye searches in order to learn our method's
strengths and weaknesses.  The only previously known blue cluster in
the survey region was G42 \cite{sargent}, and it was found by our
algorithm.  Since these regions had not been observed for young
clusters with this resolution before, all of the other coincident
cataloged objects were either individual bright stars within the
clusters or coincident emission nebulae within the confused northeast
spiral arm.  No other previously known star clusters were found. There
were, however, a few distinct clusters found within objects which had
been previously cataloged from the ground as a single cluster.  For
example, the previously cataloged H81 B-202 \cite{hodge1981}, which
was an open cluster as seen from the ground contains three of these
cluster candidates at high resolution: M31SCC J004205+405714, M31SCC
J004204+405826, and M31SCC J004205+405659.  This discovery may
indicate that many open clusters found with ground based data could in
fact be small OB associations containing multiple young clusters.  The
other clusters which were found to be part of previously cataloged
clusters are listed in Table 3.

Due to the size of the fields we were using, we were not able to find
large ($>$40 pc) OB associations previously discovered from the
ground, but we were able to identify smaller sub-clusters within these
larger associations which were not previously identified as separate
clusters.  This selection bias against large associations should be
avoidable in other data sets by looking for stellar overdensities on
larger scales, but we did not have wide enough fields to run such a
test.  We also found that our method did not detect the smallest,
densest clusters seen by eye, most likely due to the low completeness
in these areas.  It also missed several obvious red clusters,
including a few known globulars.  These clusters were either heavily
extincted or much older than the clusters found by our algorithm.  In
either case, due to severe crowding, age, or extinction, there were
too few detected blue stars in these clusters to separate the cluster
stars from the field in order to study the stellar population.

Succinctly, the algorithm tended to miss many possible clusters that
could be seen by eye.  These clusters tended to be dense red clusters
which were likely older than our sample and/or heavily reddened.
Using different color criteria, it could be possible to obtain a
sample of red cluster candidates; however, we limit our discussion
here to the blue cluster candidates which were the most
straight-forward to statistically separate from the background
population.  The clusters the algorithm did not find were too red or
too dense to easily obtain photometry of a sample of member stars.
The member stars did not stand out statistically in color, or due to
completeness, they did not stand out in stellar density.  The
algorithm found only one previously known blue cluster in the images
as well as many new clusters which were found previously by eye.  Five
of the fields had been previously scrutinized by eye looking for
extended objects which may be clusters.  Table 3 lists which of the
star clusters found by the algorithm in these five fields were also
found by eye, and which were not.  Roughly 75 percent of the objects
found by the algorithm had been previously discovered by eye on the
test fields.  The other 25 percent were comparable to M31SCC
J004455+413127 or M31SCC J004206+405649 (see Figure 5).  They did not
contain obvious compact cores and are not as likely to be real
clusters.

We also checked the overlapping fields to see which clusters were
found independently by the algorithm using photometry from different
observations of the same region.  Table 4 lists the clusters which
were in the overlapping regions of two fields, along with the fields
in which they were found.  There were clusters found in overlapping
regions of fields 9 and 10.  Only 3 of 7 cluster candidates in the
overlapping regions were found independently in both frames.  This
result reveals the dependence of our method upon the stellar density
of the field observed.  The non-overlapping regions of the overlapping
fields sample different regions, and if these non-overlapping regions
of the fields have significantly different mean stellar densities or
significantly different stellar density fluctuations, then the
algorithm will pick out some different cluster candidates for the
overlapping regions of the fields.  For example, the WF2 chip of field
10 overlaps the WF4 chip of field 9.  The non-overlapping portion of
the WF2 chip of field 10 contains a very active region.  This active
region raises the stellar density threshold for finding a cluster on
the field 10 chip.  Therefore the clusters found on the field 9 chip
are not as convincing due to the very low average stellar density,
and, in fact, these clusters are not picked out on the field 10 chip,
even though field 10 is deeper.  Exactly the inverse situation occurs
for the non-overlapping portions of the WF3 chip of fields 9 and 10.
Here the non-overlapping section of field 9 contains a very active
region, raising the threshold for finding a cluster.  This
inconsistency would likely be less severe for wider field data sets,
as the mean stellar densities and density fluctuations should be more
stable if sampled over larger regions.  The good news about these
inconsistencies between overlapping regions is that they provide a
very nice ranking of candindates in these regions.  The candidates
that were found in both data sets of the same area are much stronger
than the candidates that were not.

We show a random subset of 9 of the objects detected by the algorithm
in Figure 5.  These 9 images are through the F439W filter and are 12
arcsec across.  The example set shows a variety of objects from large
obvious clusters like M31SCC J004000+403325 to very marginal detections such
as M31SCC J004455+413127.  There were several of these types of objects in
our sample, which do not look like obvious clusters to the eye.  These
make up close to half of the sample.  Generally, the properties of
these cluster candidates are right at the limits of our criteria.  The
candidates that were not found independently when observed in two
fields were comparable to these candidates, as were the candidates
that were not seen by eye.  Nevertheless these star populations
contain well above the mean density of bright blue stars and have
enhanced surface brightness compared to the rest of the region sampled
in that field.  There was no simple objective way to remove these
objects from the sample without also removing many of our best
candidates in the process.  Though these objects are less likely bona
fide clusters, we have included them for completeness.

\section{Determining Physical Parameters for the Clusters}

\subsection{Deprojected Galactocentric Distance}

Using the coordinates of the clusters in the HST fields,
RA=00:42:44.31 DEC=41:16:09.4 for the center of M31, the inclination
angle of the disk (77 degrees) \cite{brinksandbarton1984}, and the
position angle of the disk (38 degrees), we corrected the projected
galactocentric distances of each of our fields.  First we determined the major and minor axis coordinates using:

$$
x^2 = P^2 * cos^2(\phi - \theta)
$$
$$
y^2 = (P^2 - x^2)
$$
\noindent
where x is the coordinate of the object along the major axis of M31, y
is the coordinate of the object along the minor axis, $\phi$ is the
anglular position of the object east of north, $\theta$ is the M31 position
angle east of north, and P is the projected galactocentric distance
of the field.  Then, we corrected the minor axis coordinate for the disk inclination using:

$$
y_c^2 = y^2/cos^2(I)
$$

\noindent
where $y_c$ is the minor axis coordinate corrected to its face-on value.
Together, these transformations allow a direct transformation from
P to the galactocentric distance using:

$$
G^2 = x^2 + y_c^2 = P^2 (cos^2(\phi - \theta) + (1 - cos^2(\phi - \theta))/ cos^2(I))
$$

\noindent
where G is the deprojected galactocentric distances for the clusters.
Calculated values are given in Table 2.

\subsection{Reddening and Age Determinations}

Once we had determined the positions of the clusters and the
photometry of the most likely member stars, the next step was to
correct the stellar photometry for the extinction between us and each
of the clusters.  The reddening was likely to be significant since
these clusters were all within the M31 disk where we expect there to
be relatively thick dust.  Open clusters tend to contain upper main
sequence stars.  These stars are virtually the same intrinsic color in
B-V, and they tend to lie along a narrow sequence in the U-B, B-V
plane.  Since extinction makes these stars appear to be to the red of
this sequence, it is possible to estimate the extinction values of
these clusters using photometry of just a few of the brightest cluster
members.

Since the clusters were found on the basis of the grouping of just a
handful of detected bright blue stars, we were limited in our
confidence for determining accurate reddening and age values for them.
Since there was not a statistical overdensity of red stars in the
clusters, we could not confidently assume the red stars were cluster
members.  Therefore we had very few stars from each cluster with which
to work, and it was not practical or informative to run a detailed fit
of synthetic color-magnitude diagrams to the observed color-magnitude
diagram.  With so few stars, it proved more useful and faster to
assume that the over-dense population of blue stars we detected
represented the main sequence turnoff of the cluster.  Simulations
described in section 4.3 show this to be a reasonable assumption to
estimate the age to $\sim$50\% accuracy.  This assumption allowed us
to determine reddening values by fitting model U-B and B-V star colors
to model main sequence colors when possible.  We determined the
reddening by doing a least-squares fit of the U-B and B-V colors of
the stars to the U-B and B-V colors of the theoretical main sequence
from \citeN{girardi}.  We occasionally adjusted this value in cases
where the B-V colors of the full cluster sample did not appear to
follow the isochrones due to one outlier which had polluted the
least-squares fit.  In cases where there were no U band data
available, the reddening was determined by fitting the B-V colors to
the theoretical upper main sequence.  The reddening values for the
whole sample determined by this method are given in Table 2.

Figure 6 shows a sample of one of our reddening determination fits.
The figure shows the data after applying the our best reddening
correction overplotted with the theoretical stellar colors from
\citeN{girardi}.  In some cases, only one star was detected in all
three bands due to the sensitivity of WFPC2 in the UV.  These single
star fits often had to be adjusted by eye since they often resulted in
poor fits to the B-V color for the rest of the stars detected in the
cluster candidate.  These adjustments pushed the stars slightly off
the best fit to the theoretical line in these cases.  Our findings
show that these clusters have a wide range of reddening values,
indicating that they are likely at different depths within the disk.
In the cases where U band data was not available, the reddening had to
be fit assuming that all of the stars were still on the upper main
sequence, and therefore were not red in B-V due to evolution.  Two
examples of these fits are shown in Figure 6 as well.  We assume that
all of the stars within a single cluster are reddened by the same
amount.  It is possible that some of these clusters contain dust which
would produce differential reddening within individual clusters.  If
differential reddening is affecting our data, it is possible that the
brightest stars are over-corrected.  Such an over correction would
cause an under-estimation of the cluster age as determined by the
method described below.

After correcting the photometry using these reddening values, we
produced simple least-squares fits of the stellar photometry to single
age model isochrones in order to approximate the age of the cluster.
This procedure was performed on all of our cluster candidates.  We
used all of the stars in the overdensity, without subtracting any
possible field stars.  We justify the lack of correcting for field
contamination by pointing out that the average number of these bright
blue stars in areas the size that we were sampling was generally
$\lesssim$1.  Randomly throwing out a single star from each cluster
would not have have been useful in case of our data because there was
no reason to throw out one star as opposed to another.  On the
contrary, these stars had already been selected on the basis that they
were grouped, bright blue stars of which the density in the field was
very low.  Therefore, our field contamination was minimized without
doing a second contamination correction.  All of the age fits were
inspected by eye and adjusted in cases where the fitting procedure
produced inferior fits.  The inferior fits were obvious because they
did not fit the main sequence turnoff of the cluster due to the
measured color of the brightest star in the cluster.  If this color
was measured to be bluer than the theoretical main sequence due to
photometry errors, then the least squares method could not fit a
turnoff.  In these cases, ages were determined using by-eye fits to
the isochrones.  Therefore, about half of our age determinations were
done by eye, and the fitting procedure was only used as a first
approximation.  Some typical fits are shown in Figure 7.  The ages for
the whole sample determined by this method are given in Table 2.

\subsection{Error Estimates:  Simulation Tests}

We determined the errors of our subjective age and reddening
determination technique through several experiments.  First, we
simulated our data using higher resolution PC data of well-studied
massive young clusters.  By comparing the results from our crude
method to the robust results from the high resolution data, we were
able to estimate how accurate our results were for the most populated
clusters.  Then we experimented with the clusters themselves.  By
removing data points and iterating our analysis routine, we were able
to assess the stability of our results.  Lower stability resulted in
errors larger than those determined by the resolution simulation.

As a test of our age and reddening determination method, we simulated
our low resolution wide field data using our data from the PC.  Since
we had previously obtained robust ages from the high resolution PC
data of four previously know blue globular-like clusters
\cite{Williams2001}, we binned these data 2 by 2 in order to simulate
the resolution of the WF chips ($\sim$0.1 arcsec/pixel).  We then ran the
binned data through the same photometry routine and cluster finding
algorithm as the wide field data.  This exercise was further
confirmation that the algorithm was finding clusters, as it found all
four clusters and nothing else in the frames.  Finally, we put the
stars which the algorithm provided as the most likely cluster members
through the same reddening and age determination routine in order to
check the accuracy of our method.  We found that the ages determined
were all under-estimated.  The turn-off was found at a brighter
magnitude than we had determined it from the PC. These brighter
turnoff stars were always found very near the cluster center, and
therefore were very likely blended stars at the lower resolution.
These massive and dense clusters show the worst case scenario for this
kind of blending, so that we hope our less populous clusters will not
suffer as badly from this effect.

A summary of the results of the low resolution simulation is shown in
Table 5.  The ages are systematically under estimated by a mean of 0.2
dex.  Since the vast majority of the new cluster candidates were not
this dense and massive, we did not feel that it would be appropriate
to simply add 0.2 dex to all of our ages.  Rather, this result was
useful for determining our error bars.  The test showed that $\pm$0.2
dex was the best we could do for our age determinations.  The test
result also showed that error determination by throwing out the
brightest few stars at the turnoff and redetermining the reddening and
age was reasonable.

In order to assess the errors of our reddening and age estimates for
each cluster candidate individually, we removed the brightest star
from each small cluster and the brightest four stars from every large
cluster.  We then measured the reddening and age again using the same
method.  This experiment allowed us to find the stability of our
estimate as well as to account for the effect of field stars, and/or
blends on our results.  The errors given are the difference between
the reddening and age values determined with the full sample of stars
and the values determined with the manipulated sample.  If this error
value was less than 0.1 in $E_{B-V}$ or 0.2 in log age, the error was
set to 0.1 in $E_{B-V}$ or 0.2 in log age because our low resolution
simulation had shown that our errors had to be at least this large.
We therefore would not allow these reddening and age values,
determined using a much smaller number of stars, to have smaller
quoted errors.  These small experimental errors were quite common,
mostly due to the high sensitivity of the turnoff to the age at these
young ages (2 mag between 30 Myr and 100 Myr).  It was encouraging to
find that most of our experiments showed our results to be quite
stable against removal of the turnoff data points.

\section{Results and Conclusions}

All of our measurements are given in Table 2.  With this table, we
have provided an objective collection of young star cluster candidates
in the M31 disk along with their reddenings and ages as determined
from the photometry of their constituent stars.  With our crude age
approximations, it was possible to look for statistical patterns in
the age distribution.  While the precision of our ages is clearly low,
and in fact we cannot even quote reliable errors on our age
measurements, our data is most sensitive to bright blue stars which
define the main sequence turnoff of these clusters.  Assuming the
detected stars do mark the turnoff, and that our data set was equally
sensitive to them for all of the clusters detected, we believe that we
have done the best job possible to preserve their relative ages by
reducing all of the data identically and determining all ages with the
same model isochrones.

We looked for correlations between the reddening of the clusters and
their galactocentric distances and between the ages of the clusters
and their galactocentric distances.  These plots are shown in Figure
8.  No correlation is seen within the large errors in our
measurements.  Apparently, more accurate age determinations are needed
in order to dicipher the propagation of recent cluster formation in
these regions of the M31 disk.  We also checked for a correlation
between ages and reddening which would have been indicative of
observational biases within the sample.  As seen in Figure 9, there
appears to be a lack of older clusters with high reddening values,
confirming that our selection method is bias against such clusters.
There is also a lack of older clusters with very low reddening.  This
effect was not expected, and it may be an effect of field
contaminants.  It is possible that the six clusters with very low
reddening have field contaminants causing them to appear less reddened
and younger.

We have created an automated routine for finding young star clusters
amoung stellar populations in nearby galaxies.  This method requires
the clusters to be resolved into individual stars, so that the
positions and photometric properties of the stars can be used to
distinguish the star cluster candidate from the field.  From
comparisons between overlapping data sets and comparisons between the
automated routine and independent searches by eye, we expect at least
half of these candidates are real, young star clusters.  The method is
not effective for finding small compact clusters whose stellar
populations cannot be studied in detail with the survey data.
Unfortunately, the algorithm misses clusters that could be useful when
higher resolution data are obtainable, and the algorithm finds some
very loose associations which are likely not real clusters.  On the
other hand, the method is quite effective at finding star clusters
whose populations can be further studied with the survey data.  The
algorithm only finds clusters whose stellar photometry is
statistically different from the surrounding stellar population.  This
statistical difference provides a sample of stars from each cluster
candidate which can be used to constrain the age and reddening.

We have objectively detected 80 blue cluster candidates in the M31
disk using HST/WFPC2 archival data of 13 fields.  Of these clusters,
79 are newly discovered as individual clusters, though many lie within
previously known OB associations.  We have determined rough ages and
extinction estimates from the stellar photometry.  The ages and
reddening values for these clusters span the full range of our
sensitivity and are consistent with the range of ages and reddening
values of the OB associations determined by \citeN{magnier1997} in the
fields common to both studies.  The precision of our approximations is
too low to look for cluster formation patterns; however, future, more
accurate determinations of these values could significantly advance
our understanding of the propagation of cluster formation in the M31
disk.

\section{Acknowledgments}

Support for this work was provided by NASA through grant number
GO-06459.01-95A from the Space Telescope Science Institute, which is
operated by the Association of Universities for Research in Astronomy,
Incorporated, under NASA contract NAS5-26555.



\clearpage

\begin{deluxetable}{ccccccc}
\small
\tablecaption{Data obtained from the HST data archive used for the cluster survey.}
\tableheadfrac{0.05}
\tablehead{
\colhead{\bf{Field}} &
\colhead{\bf{Prop. \#}} &
\colhead{\bf{Obs. date}} &
\colhead{\bf{RA (2000)}} &
\colhead{\bf{DEC (2000)}} &
\colhead{\bf{Filter}}  &
\colhead{\bf{Exp. (sec)}} 
}
\startdata
1 & 8296 & Oct 15 1999 & 0:39:47.35 & 40:31:57.9 & F336W & 1000\nl
1 & 8296 & Oct 15 1999 & 0:39:47.35 & 40:31:57.9 & F336W & 800\nl
1 & 8296 & Oct 15 1999 & 0:39:47.35 & 40:31:57.9 & F336W & 1200\nl
1 & 8296 & Oct 15 1999 & 0:39:47.35 & 40:31:57.9 & F336W & 600\nl
1 & 8296 & Oct 15 1999 & 0:39:47.35 & 40:31:57.9 & F439W & 800\nl
1 & 8296 & Oct 15 1999 & 0:39:47.35 & 40:31:57.9 & F439W & 800\nl
1 & 8296 & Oct 15 1999 & 0:39:47.35 & 40:31:57.9 & F555W & 600\nl
1 & 8296 & Oct 15 1999 & 0:39:47.35 & 40:31:57.9 & F555W & 600\nl
2 & 8296 & Oct 30 1999 & 0:40:01.58 & 40:34:14.7 & F336W & 1000\nl
2 & 8296 & Oct 30 1999 & 0:40:01.58 & 40:34:14.7 & F336W & 800\nl
2 & 8296 & Oct 30 1999 & 0:40:01.58 & 40:34:14.7 & F336W & 1200\nl
2 & 8296 & Oct 30 1999 & 0:40:01.58 & 40:34:14.7 & F336W & 600\nl
2 & 8296 & Oct 30 1999 & 0:40:01.58 & 40:34:14.7 & F439W & 800\nl
2 & 8296 & Oct 30 1999 & 0:40:01.58 & 40:34:14.7 & F439W & 800\nl
2 & 8296 & Oct 30 1999 & 0:40:01.58 & 40:34:14.7 & F555W & 600\nl
2 & 8296 & Oct 30 1999 & 0:40:01.58 & 40:34:14.7 & F555W & 600\nl
3 & 6038 & Jan 23 1996 & 0:40:14.10 & 40:37:11.3 & F336W & 900\nl
3 & 6038 & Jan 23 1996 & 0:40:14.10 & 40:37:11.3 & F336W & 900\nl
3 & 6038 & Jan 23 1996 & 0:40:14.10 & 40:37:11.3 & F439W & 600\nl
3 & 6038 & Jan 23 1996 & 0:40:14.10 & 40:37:11.3 & F555W & 160\nl
4 & 6431 & Dec 9 1997 & 0:40:39.54 & 40:33:25.4 & F439W & 350\nl
4 & 6431 & Dec 9 1997 & 0:40:39.54 & 40:33:25.4 & F439W & 350\nl
4 & 6431 & Dec 9 1997 & 0:40:39.54 & 40:33:25.4 & F555W & 260\nl
4 & 6431 & Dec 9 1997 & 0:40:39.54 & 40:33:25.4 & F555W & 260\nl
4 & 6431 & Dec 9 1997 & 0:40:39.54 & 40:33:25.4 & F814W & 260\nl
4 & 6431 & Dec 9 1997 & 0:40:39.54 & 40:33:25.4 & F814W & 260\nl
5 & 8296 & Oct 30 1999 & 0:41:22.08 & 40:37:06.7 & F336W & 600\nl
5 & 8296 & Oct 30 1999 & 0:41:22.08 & 40:37:06.7 & F336W & 1000\nl
5 & 8296 & Oct 30 1999 & 0:41:22.08 & 40:37:06.7 & F336W & 800\nl
5 & 8296 & Oct 30 1999 & 0:41:22.08 & 40:37:06.7 & F336W & 1200\nl
5 & 8296 & Oct 30 1999 & 0:41:22.08 & 40:37:06.7 & F439W & 800\nl
5 & 8296 & Oct 30 1999 & 0:41:22.08 & 40:37:06.7 & F439W & 800\nl
5 & 8296 & Oct 30 1999 & 0:41:22.08 & 40:37:06.7 & F555W & 600\nl
5 & 8296 & Oct 30 1999 & 0:41:22.08 & 40:37:06.7 & F555W & 600\tablebreak
6 & 6431 & Dec 9 1997 & 0:42:05.27 & 40:57:33.9 & F439W & 350\nl
6 & 6431 & Dec 9 1997 & 0:42:05.27 & 40:57:33.9 & F439W & 350\nl
6 & 6431 & Dec 9 1997 & 0:42:05.27 & 40:57:33.9 & F555W & 260\nl
6 & 6431 & Dec 9 1997 & 0:42:05.27 & 40:57:33.9 & F555W & 260\nl
6 & 6431 & Dec 9 1997 & 0:42:05.27 & 40:57:33.9 & F814W & 260\nl
6 & 6431 & Dec 9 1997 & 0:42:05.27 & 40:57:33.9 & F814W & 260\nl
7 & 5911 & Oct 3 1995 & 0:44:44.17 & 41:27:33.8 & F336W & 400\nl
7 & 5911 & Oct 3 1995 & 0:44:44.23 & 41:27:33.8 & F439W & 160\nl
7 & 5911 & Oct 3 1995 & 0:44:44.23 & 41:27:33.8 & F555W & 140\nl
8 & 8296 & Oct 31 1999 & 0:44:46.19 & 41:51:33.3 & F336W & 1000\nl
8 & 8296 & Oct 31 1999 & 0:44:46.19 & 41:51:33.3 & F336W & 800\nl
8 & 8296 & Oct 31 1999 & 0:44:46.19 & 41:51:33.3 & F336W & 1200\nl
8 & 8296 & Oct 31 1999 & 0:44:46.19 & 41:51:33.3 & F336W & 600\nl
8 & 8296 & Oct 31 1999 & 0:44:46.19 & 41:51:33.3 & F439W & 800\nl
8 & 8296 & Oct 31 1999 & 0:44:46.19 & 41:51:33.3 & F439W & 800\nl
8 & 8296 & Oct 31 1999 & 0:44:46.19 & 41:51:33.3 & F555W & 600\nl
8 & 8296 & Oct 31 1999 & 0:44:46.19 & 41:51:33.3 & F555W & 600\nl
9 & 5911 & Oct 8 1995 & 0:44:49.28 & 41:28:59.0 & F336W & 400\nl
9 & 5911 & Oct 8 1995 & 0:44:49.34 & 41:28:59.0 & F439W & 160\nl
9 & 5911 & Oct 8 1995 & 0:44:49.34 & 41:28:59.0 & F555W & 140\nl
10 & 6038 & Jan 1 1996 & 0:44:51.22 & 41:30:03.7 & F336W & 900\nl
10 & 6038 & Jan 1 1996 & 0:44:51.22 & 41:30:03.7 & F336W & 900\nl
10 & 6038 & Jan 1 1996 & 0:44:51.22 & 41:30:03.7 & F439W & 600\nl
10 & 6038 & Jan 1 1996 & 0:44:51.22 & 41:30:03.7 & F555W & 160\nl
11 & 5911 & Oct 4 1995 & 0:44:57.57 & 41:30:51.6 & F336W & 400\nl
11 & 5911 & Oct 4 1995 & 0:44:57.63 & 41:30:51.6 & F439W & 160\nl
11 & 5911 & Oct 4 1995 & 0:44:57.63 & 41:30:51.6 & F555W & 140\nl
12 & 5911 & Oct 15 1995 & 0:45:09.20 & 41:34:30.5 & F336W & 400\nl
12 & 5911 & Oct 15 1995 & 0:45:09.25 & 41:34:30.7 & F439W & 160\nl
12 & 5911 & Oct 15 1995 & 0:45:09.25 & 41:34:30.7 & F555W & 140\nl
13 & 5911 & Oct 15 1995 & 0:45:11.89 & 41:36:56.8 & F336W & 400\nl
13 & 5911 & Oct 15 1995 & 0:45:11.95 & 41:36:57.0 & F439W & 160\nl
13 & 5911 & Oct 15 1995 & 0:45:11.95 & 41:36:57.0 & F555W & 140\nl
\hline
\enddata
\end{deluxetable}

\clearpage
\begin{landscape}
\oddsidemargin -1.5cm
\begin{deluxetable}{ccccccc}
\tablenotetext{a}{M31SCC is an IAU registered acronym; G42 refers to the globular cluster catalog of \citeN{sargent}; [H81] B-202 is identified in \citeN{hodge1981}}
\tablenotetext{b}{Search radius given to the automated search routine.  Using this radius the algorithm searched for overdensities over areas of ($\pi$$R_s^2$).}
\small
\tablecaption{Catalog of positions, galactocentric distances, age estimates, reddening values, and search radii for M31 young star cluster candidates.}
\tableheadfrac{0.05}
\tablewidth{8.5in}
\tablehead{
\colhead{{ID\tablenotemark{a}}} &
\colhead{{RA (2000)}} &
\colhead{{DEC (2000)}} &
\colhead{{GCD (kpc)}} &
\colhead{{log AGE}}  &
\colhead{{${E_{B-V}}$}} &
\colhead{{$R\tablenotemark{b}_{s}$(pc)}} 
}

\oddsidemargin -1.5cm
\startdata
\oddsidemargin -1.5cm
M31SCC J003952+403141 &  0:39:52.43 & 40:31:41.27 & 15.21 & 8.00$\pm$0.20 & 0.28$\pm$0.10 & 5\nl
M31SCC J004000+403325 &  0:40:00.03 & 40:33:25.02 & 14.52 & 7.90$\pm$0.35 & 0.22$\pm$0.10 & 15\nl
M31SCC J004000+403406 (G42) &  0:40:00.83 & 40:34:06.64 & 14.49 & 7.75$\pm$0.45 & 0.21$\pm$0.10 & 15\nl
M31SCC J004001+403420 &  0:40:01.54 & 40:34:20.06 & 14.44 & 8.25$\pm$0.30 & 0.20$\pm$0.20 & 5\nl
M31SCC J004004+403440 &  0:40:04.66 & 40:34:40.51 & 14.11 & 8.10$\pm$0.20 & 0.23$\pm$0.10 & 5\nl
M31SCC J004006+403508 &  0:40:06.78 & 40:35:08.41 & 13.92 & 7.30$\pm$0.85 & 0.25$\pm$0.10 & 15\nl
M31SCC J004010+403624 &  0:40:10.36 & 40:36:24.16 & 13.63 & 7.75$\pm$0.60 & 0.01$\pm$0.10 & 5\nl
M31SCC J004012+403632 &  0:40:12.01 & 40:36:32.62 & 13.46 & 7.70$\pm$0.50 & 0.24$\pm$0.10 & 5\nl
M31SCC J004012+403617 &  0:40:12.97 & 40:36:17.32 & 13.33 & 7.90$\pm$0.45 & 0.17$\pm$0.10 & 15\nl
M31SCC J004013+403815 &  0:40:13.68 & 40:38:15.54 & 13.46 & 7.70$\pm$0.40 & 0.43$\pm$0.10 & 5\nl
M31SCC J004015+403652 &  0:40:15.42 & 40:36:52.85 & 13.11 & 7.90$\pm$0.20 & 0.35$\pm$0.10 & 5\nl
M31SCC J004032+403320 &  0:40:32.93 & 40:33:20.45 & 12.10 & 7.90$\pm$0.20 & 0.16$\pm$0.10 & 5\nl
M31SCC J004033+403308a &  0:40:33.28 & 40:33:08.64 & 12.13 & 7.75$\pm$0.20 & 0.31$\pm$0.10 & 5\nl
M31SCC J004033+403326 &  0:40:33.46 & 40:33:26.93 & 12.06 & 7.85$\pm$0.20 & 0.30$\pm$0.10 & 5\nl
M31SCC J004033+403319 &  0:40:33.48 & 40:33:19.15 & 12.09 & 7.75$\pm$0.40 & 0.30$\pm$0.10 & 5\nl
M31SCC J004033+403308b &  0:40:33.51 & 40:33:08.39 & 12.12 & 8.00$\pm$0.20 & 0.24$\pm$0.10 & 5\nl
M31SCC J004033+403346 &  0:40:33.70 & 40:33:46.69 & 11.99 & 8.00$\pm$0.20 & 0.34$\pm$0.10 & 5\nl
M31SCC J004034+403351 &  0:40:34.62 & 40:33:51.59 & 11.95 & 8.00$\pm$0.20 & 0.32$\pm$0.10 & 15\nl
M31SCC J004035+403420 &  0:40:35.45 & 40:34:20.24 & 11.83 & 7.90$\pm$0.20 & 0.28$\pm$0.10 & 5\nl
M31SCC J004035+403251 &  0:40:35.51 & 40:32:51.18 & 12.15 & 7.85$\pm$0.20 & 0.22$\pm$0.10 & 5\nl
M31SCC J004039+403210 &  0:40:39.35 & 40:32:10.79 & 12.32 & 8.15$\pm$0.35 & 0.31$\pm$0.18 & 5\nl
M31SCC J004040+403223 &  0:40:40.43 & 40:32:23.50 & 12.26 & 7.85$\pm$0.20 & 0.27$\pm$0.10 & 5\tablebreak
M31SCC J004040+403256 &  0:40:40.67 & 40:32:56.98 & 12.10 & 8.10$\pm$0.40 & 0.37$\pm$0.10 & 5\nl
M31SCC J004041+403222 &  0:40:41.35 & 40:32:22.88 & 12.27 & 7.90$\pm$0.30 & 0.22$\pm$0.10 & 5\nl
M31SCC J004117+403720 &  0:41:17.21 & 40:37:20.57 & 11.92 & 7.85$\pm$0.20 & 0.34$\pm$0.11 & 5\nl
M31SCC J004119+403748 &  0:41:19.57 & 40:37:48.79 & 11.89 & 8.10$\pm$0.40 & 0.27$\pm$0.13 & 5\nl
M31SCC J004121+403638 &  0:41:21.37 & 40:36:38.84 & 12.65 & 7.45$\pm$0.40 & 0.45$\pm$0.10 & 5\nl
M31SCC J004123+403726 &  0:41:23.89 & 40:37:26.98 & 12.47 & 7.85$\pm$0.70 & 0.02$\pm$0.10 & 5\nl
M31SCC J004125+403723 &  0:41:25.92 & 40:37:23.66 & 12.71 & 8.20$\pm$0.45 & 0.29$\pm$0.28 & 5\nl
M31SCC J004158+405738 &  0:41:58.57 & 40:57:38.48 &  5.40 & 7.80$\pm$0.30 & 0.50$\pm$0.10 & 5\nl
M31SCC J004204+405826 (H81 B-202) &  0:42:04.85 & 40:58:26.44 &  5.47 & 7.95$\pm$0.20 & 0.43$\pm$0.11 & 5\nl
M31SCC J004205+405714 (H81 B-202) &  0:42:05.59 & 40:57:14.26 &  6.15 & 7.30$\pm$0.20 & 0.22$\pm$0.10 & 5\nl
M31SCC J004205+405659 (H81 B-202) &  0:42:05.80 & 40:56:59.06 &  6.31 & 7.75$\pm$0.45 & 0.37$\pm$0.10 & 5\nl
M31SCC J004206+405649 &  0:42:06.23 & 40:56:49.38 &  6.45 & 7.40$\pm$0.45 & 0.35$\pm$0.10 & 5\nl
M31SCC J004207+405801 &  0:42:07.11 & 40:58:01.27 &  5.89 & 7.25$\pm$0.55 & 0.35$\pm$0.10 & 15\nl
M31SCC J004441+412701 &  0:44:41.19 & 41:27:01.37 & 17.32 & 7.45$\pm$0.20 & 0.38$\pm$0.10 & 5\nl
M31SCC J004441+415136 &  0:44:41.35 & 41:51:36.86 & 10.37 & 8.05$\pm$0.20 & 0.24$\pm$0.10 & 5\nl
M31SCC J004441+415239 &  0:44:41.61 & 41:52:39.22 & 10.50 & 7.75$\pm$0.30 & 0.29$\pm$0.10 & 5\nl
M31SCC J004441+415123 &  0:44:41.79 & 41:51:23.15 & 10.37 & 8.05$\pm$0.20 & 0.41$\pm$0.10 & 5\nl
M31SCC J004442+415237 &  0:44:42.09 & 41:52:37.45 & 10.52 & 7.95$\pm$0.20 & 0.34$\pm$0.10 & 5\nl
M31SCC J004442+415153 &  0:44:42.48 & 41:51:53.42 & 10.46 & 7.85$\pm$0.20 & 0.35$\pm$0.10 & 5\nl
M31SCC J004444+412749 &  0:44:44.86 & 41:27:49.32 & 17.63 & 7.40$\pm$0.60 & 0.22$\pm$0.10 & 5\nl
M31SCC J004445+412800 &  0:44:45.07 & 41:28: 0.48 & 17.58 & 7.95$\pm$0.20 & 0.41$\pm$0.14 & 5\nl
M31SCC J004445+415121 &  0:44:45.39 & 41:51:21.42 & 10.61 & 7.75$\pm$0.20 & 0.35$\pm$0.10 & 5\tablebreak
M31SCC J004445+415208 &  0:44:45.68 & 41:52:08.65 & 10.68 & 8.25$\pm$0.20 & 0.30$\pm$0.10 & 5\nl
M31SCC J004445+415107 &  0:44:45.73 & 41:51:07.88 & 10.63 & 7.50$\pm$0.45 & 0.26$\pm$0.10 & 5\nl
M31SCC J004447+415238 &  0:44:47.06 & 41:52:38.21 & 10.81 & 8.05$\pm$0.20 & 0.42$\pm$0.10 & 5\nl
M31SCC J004447+412821 &  0:44:47.32 & 41:28:21.97 & 17.84 & 7.70$\pm$0.20 & 0.02$\pm$0.10 & 5\nl
M31SCC J004447+412843 &  0:44:47.74 & 41:28:43.82 & 17.73 & 7.30$\pm$0.35 & 0.27$\pm$0.10 & 15\nl
M31SCC J004448+412925 &  0:44:48.43 & 41:29:25.15 & 17.52 & 7.75$\pm$0.20 & 0.02$\pm$0.10 & 5\nl
M31SCC J004449+413034 &  0:44:49.04 & 41:30:34.78 & 17.07 & 7.55$\pm$0.20 & 0.55$\pm$0.10 & 5\nl
M31SCC J004449+412924 &  0:44:49.25 & 41:29:24.97 & 17.68 & 7.50$\pm$0.50 & 0.23$\pm$0.10 & 5\nl
M31SCC J004449+415131 &  0:44:49.86 & 41:51:31.39 & 10.97 & 7.70$\pm$0.20 & 0.42$\pm$0.10 & 5\nl
M31SCC J004450+415211 &  0:44:50.26 & 41:52:11.50 & 11.00 & 8.15$\pm$0.30 & 0.18$\pm$0.10 & 5\nl
M31SCC J004450+412917 &  0:44:50.38 & 41:29:17.95 & 17.96 & 7.60$\pm$0.35 & 0.29$\pm$0.10 & 5\nl
M31SCC J004450+412914 &  0:44:50.97 & 41:29:14.60 & 18.11 & 7.50$\pm$0.20 & 0.16$\pm$0.10 & 5\nl
M31SCC J004451+412924 &  0:44:51.32 & 41:29:24.83 & 18.09 & 7.50$\pm$0.50 & 0.02$\pm$0.10 & 5\nl
M31SCC J004451+412911 &  0:44:51.74 & 41:29:11.44 & 18.28 & 7.35$\pm$0.40 & 0.13$\pm$0.10 & 15\nl
M31SCC J004452+415144 &  0:44:52.35 & 41:51:44.21 & 11.19 & 7.80$\pm$0.25 & 0.23$\pm$0.10 & 5\nl
M31SCC J004453+412927 &  0:44:53.52 & 41:29:27.82 & 18.49 & 7.80$\pm$0.30 & 0.21$\pm$0.10 & 5\nl
M31SCC J004455+413127 &  0:44:55.90 & 41:31:27.16 & 17.96 & 7.65$\pm$0.20 & 0.24$\pm$0.10 & 5\nl
M31SCC J004456+413121 &  0:44:56.02 & 41:31:21.54 & 18.03 & 7.35$\pm$0.20 & 0.21$\pm$0.10 & 5\nl
M31SCC J004457+413123 &  0:44:57.73 & 41:31:23.30 & 18.35 & 7.70$\pm$0.20 & 0.01$\pm$0.10 & 5\nl
M31SCC J004458+413049 &  0:44:58.97 & 41:30:49.21 & 18.87 & 7.65$\pm$0.40 & 0.33$\pm$0.10 & 5\nl
M31SCC J004500+413057 &  0:45:00.93 & 41:30:57.71 & 19.18 & 7.30$\pm$0.40 & 0.41$\pm$0.10 & 15\nl
M31SCC J004503+413408 &  0:45:03.82 & 41:34:08.54 & 18.19 & 7.90$\pm$0.25 & 0.20$\pm$0.10 & 5\tablebreak
M31SCC J004504+413451 &  0:45:04.52 & 41:34:51.60 & 17.99 & 7.95$\pm$0.20 & 0.25$\pm$0.30 & 5\nl
M31SCC J004506+413406 &  0:45:06.32 & 41:34:06.96 & 18.68 & 7.70$\pm$0.20 & 0.45$\pm$0.10 & 5\nl
M31SCC J004506+413545 &  0:45:06.73 & 41:35:45.78 & 17.99 & 8.00$\pm$0.50 & 0.21$\pm$0.20 & 15\nl
M31SCC J004509+413643 &  0:45:09.82 & 41:36:43.31 & 18.14 & 7.80$\pm$0.20 & 0.10$\pm$0.10 & 5\nl
M31SCC J004509+413649 &  0:45:09.82 & 41:36:49.79 & 18.09 & 7.30$\pm$0.50 & 0.31$\pm$0.10 & 5\nl
M31SCC J004510+413645 &  0:45:10.33 & 41:36:45.47 & 18.22 & 7.25$\pm$0.20 & 0.14$\pm$0.10 & 5\nl
M31SCC J004511+413711 &  0:45:11.82 & 41:37:11.86 & 18.30 & 7.90$\pm$0.30 & 0.24$\pm$0.10 & 5\nl
M31SCC J004512+413715 &  0:45:12.31 & 41:37:15.78 & 18.36 & 7.95$\pm$0.20 & 0.11$\pm$0.10 & 5\nl
M31SCC J004512+413716 &  0:45:12.48 & 41:37:16.82 & 18.38 & 7.70$\pm$0.40 & 0.33$\pm$0.10 & 5\nl
M31SCC J004512+413723 &  0:45:12.78 & 41:37:23.45 & 18.39 & 8.00$\pm$0.50 & 0.23$\pm$0.10 & 5\nl
M31SCC J004512+413727 &  0:45:12.87 & 41:37:27.80 & 18.38 & 7.80$\pm$0.20 & 0.25$\pm$0.10 & 5\nl
M31SCC J004513+413735 &  0:45:13.40 & 41:37:35.08 & 18.42 & 7.30$\pm$0.20 & 0.46$\pm$0.10 & 5\nl
M31SCC J004514+413743 &  0:45:14.13 & 41:37:43.72 & 18.49 & 7.35$\pm$0.20 & 0.39$\pm$0.10 & 5\nl
M31SCC J004514+413724 &  0:45:14.22 & 41:37:24.31 & 18.66 & 7.70$\pm$0.20 & 0.36$\pm$0.10 & 5\nl
\hline
\enddata
\end{deluxetable}
\topmargin 0.0cm
\end{landscape}
\clearpage
\oddsidemargin 0.0cm

\begin{deluxetable}{cccc} 
\tablenotetext{a}{M31SCC is an IAU registered acronym; 
the [H81] B prefix is identified in \citeN{hodge1981} table B.}
\small
\tablecolumns{4}
\tablecaption{Comparison of 5 fields searched both by eye and by algorithm.  About 75 percent of the cluster candidates were found by both methods.  Many of the candidates are parts of previously known associations.} 
\tablehead{
\colhead{\bf{Field}} &
\colhead{\bf{Cluster}} &
\colhead{\bf{Found by eye?}} &
\colhead{\bf{Association\tablenotemark{a}}}
}                                  
\startdata
7 & M31SCC J004441+412701 & N &  \nl
7 & M31SCC J004444+412749 & Y &  \nl
7 & M31SCC J004445+412800 & N &  \nl
7 & M31SCC J004447+412821 & Y &  \nl
9 & M31SCC J004447+412843 & Y & Part of [H81] B-298 \nl
9 & M31SCC J004451+412911 & Y & Part of [H81] B-301 \nl
9 & M31SCC J004453+412927 & N &  \nl
11 & M31SCC J004455+413127 & Y & Part of [H81] B-306 \nl
11 & M31SCC J004456+413121 & Y & Part of [H81] B-306 \nl
11 & M31SCC J004457+413123 & Y &  \nl
11 & M31SCC J004458+413049 & Y & \nl 
11 & M31SCC J004500+413057 & Y &  \nl
12 & M31SCC J004503+413408 & Y &  \nl
12 & M31SCC J004504+413451 & N &  \nl
12 & M31SCC J004506+413406 & N &  \nl
12 & M31SCC J004506+413545 & N &  \nl
13 & M31SCC J004509+413643 & Y & Part of [H81] B-310 \nl
13 & M31SCC J004509+413649 & Y & Part of [H81] B-310 \nl
13 & M31SCC J004510+413645 & Y & Part of [H81] B-310 \nl
13 & M31SCC J004511+413711 & Y & \nl 
13 & M31SCC J004512+413715 & Y & Part of [H81] B-312 \nl
13 & M31SCC J004512+413716 & Y & Part of [H81] B-312 \nl
13 & M31SCC J004512+413723 & N & Part of [H81] B-312 \nl
13 & M31SCC J004512+413727 & Y &  \nl
13 & M31SCC J004513+413735 & Y & \nl
13 & M31SCC J004514+413743 & Y & \nl
13 & M31SCC J004514+413724 & Y & \nl
\hline
\enddata
\end{deluxetable}

\begin{deluxetable}{cccc} 
\small
\tablecolumns{3}
\tablecaption{Comparison of detection cluster candidates appearing in more than one field.} 
\tablehead{
\colhead{\bf{Cluster}} &
\colhead{\bf{Fields Observed}} &
\colhead{\bf{Fields Found}}
}                                  
\startdata
M31SCC J004448+412925 & 9,10 & 10 \nl
M31SCC J004449+412924 & 9,10 & 10 \nl
M31SCC J004450+412914 & 9,10 & 9,10 \nl
M31SCC J004450+412917 & 9,10 & 9,10 \nl
M31SCC J004451+412924 & 9,10 & 9,10 \nl
M31SCC J004451+412911 & 9,10 & 9 \nl
M31SCC J004453+412927 & 9,10 & 9 \nl
\hline
\enddata
\end{deluxetable}

\begin{deluxetable}{ccccc} 
\small
\tablecolumns{5}
\tablecaption{Comparison of ages determined by isochrone fitting on high and low resolution WFPC2 photometry.} 
\tablehead{
\colhead{\bf{Name}} &
\colhead{\bf{log Age$_{\bf{hires}}$ (yr)}} &
\colhead{\bf{\bf{log Age$_{\bf{lowres}}$ (yr)}}} &
\colhead{\bf{\bf{E(B-V)$_{\bf{hires}}$}}} &
\colhead{\bf{\bf{E(B-V)$_{\bf{lowres}}$}}}
}                                  
\startdata
G38 & 8.00$\pm$0.15 & 7.75$\pm$0.20 & 0.31$\pm$0.11 & 0.26$\pm$0.10\nl
G44 & 8.00$\pm$0.15 & 7.90$\pm$0.20 & 0.23$\pm$0.10 & 0.21$\pm$0.10\nl
G94 & 8.20$\pm$0.15 & 7.80$\pm$0.20 & 0.20$\pm$0.10 & 0.39$\pm$0.10\nl
G293 & 7.80$\pm$0.10 & 7.65$\pm$0.20 & 0.20$\pm$0.10 & 0.35$\pm$0.10\nl
\hline
\enddata
\end{deluxetable}

\clearpage

\begin{figure} 
\centerline{\psfig{file=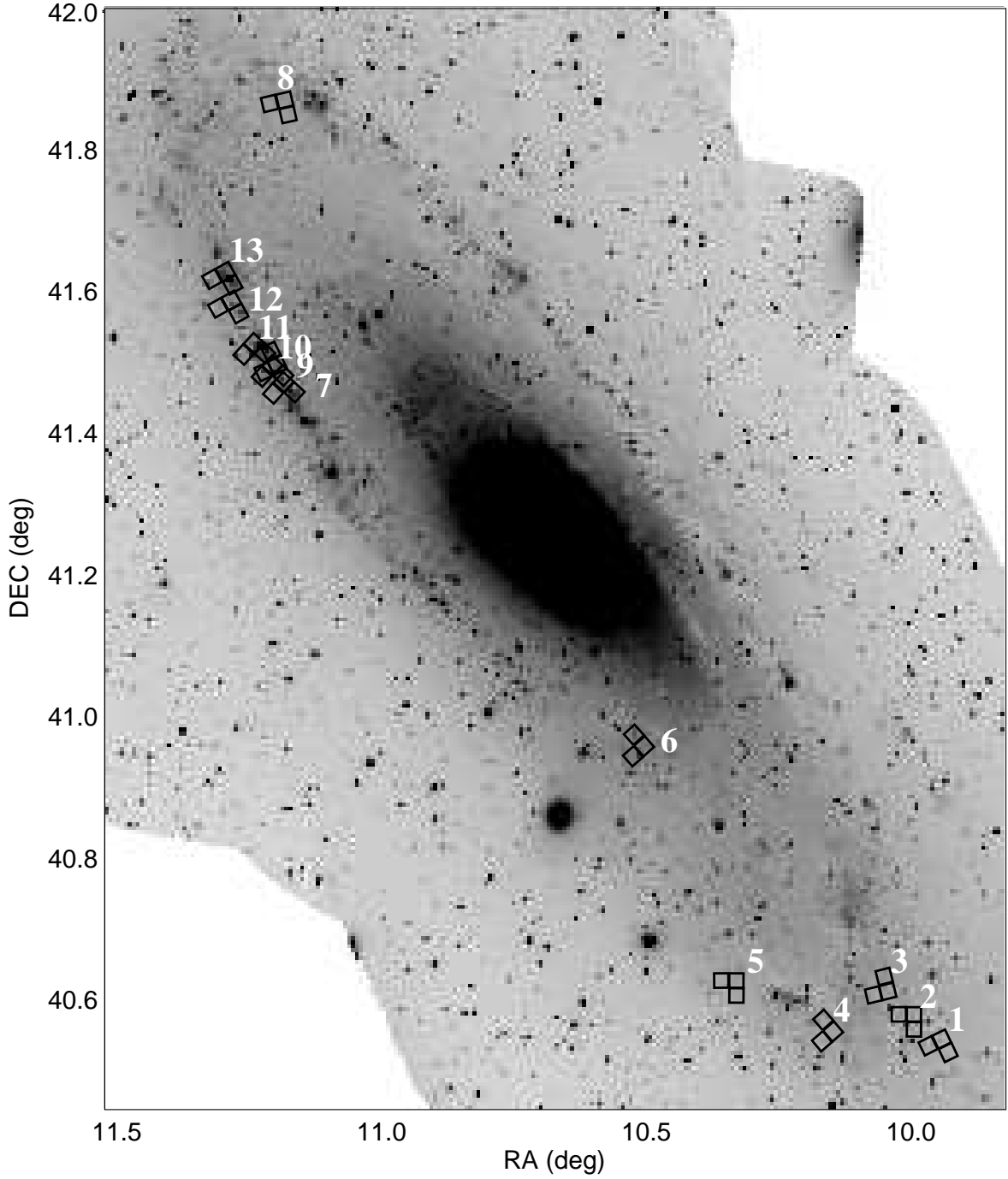 ,height=8.0in,angle=270}}
\caption{Positions of the HST fields taken from the HST archive.  Shown are the fields that were observed through blue filters allowing detailed studies of the young main sequence population.} 
\end{figure}

\begin{figure} 
\centerline{\psfig{file=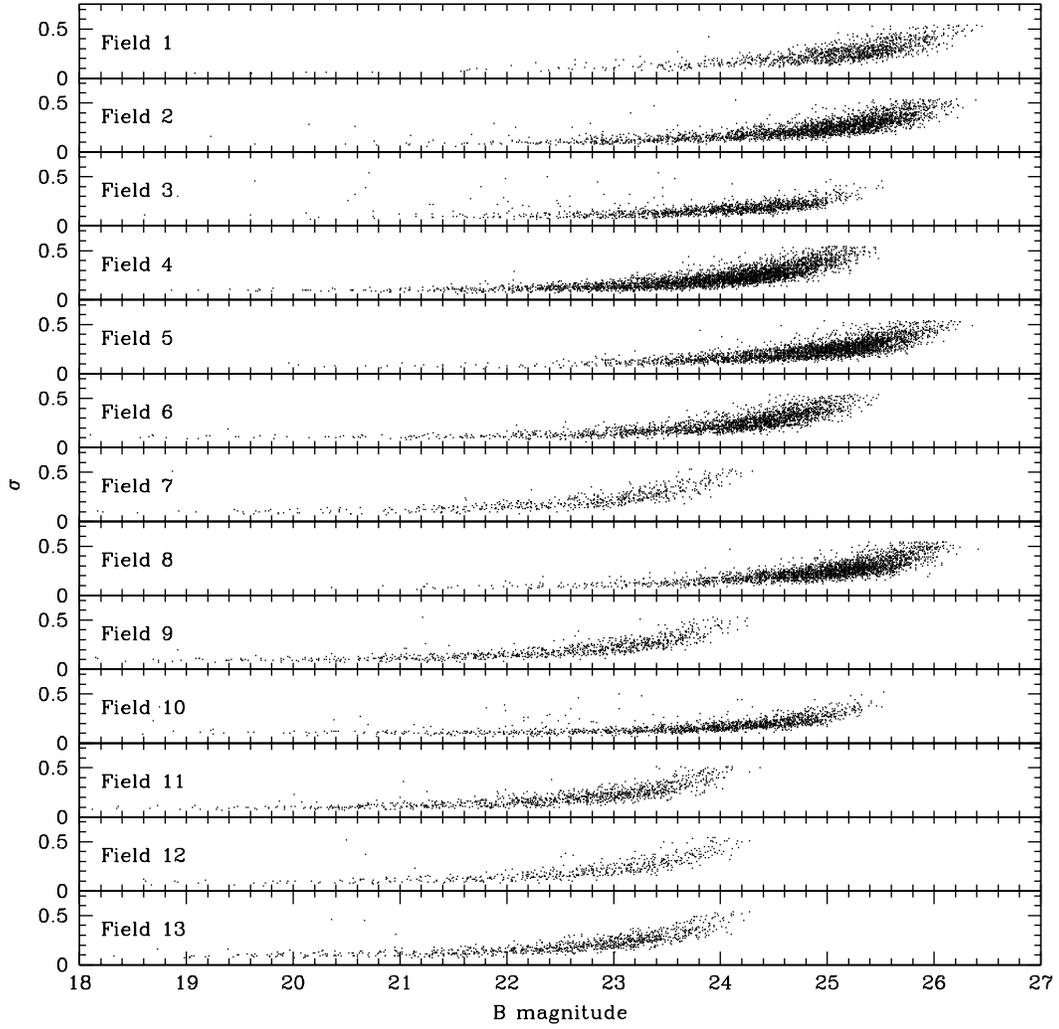,height=6.0in,angle=0}}
\caption{B band (F439W) photometric errors from ALLSTAR for the 13 fields.} 
\end{figure}

\begin{figure} 
\centerline{\psfig{file=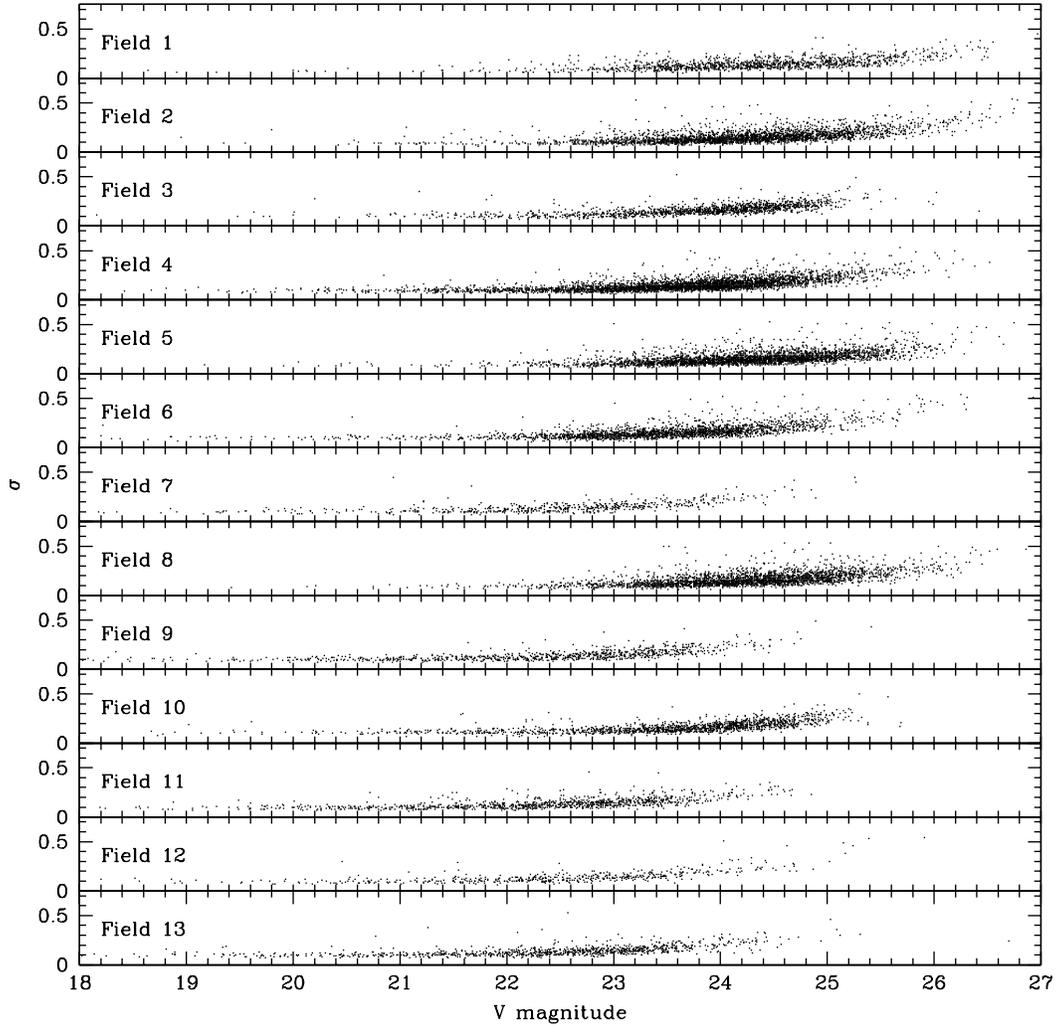,height=6.0in,angle=0}}
\caption{V band (F555W) photometric errors from ALLSTAR for the 13 fields.} 
\end{figure}

\begin{figure} 
\centerline{\psfig{file=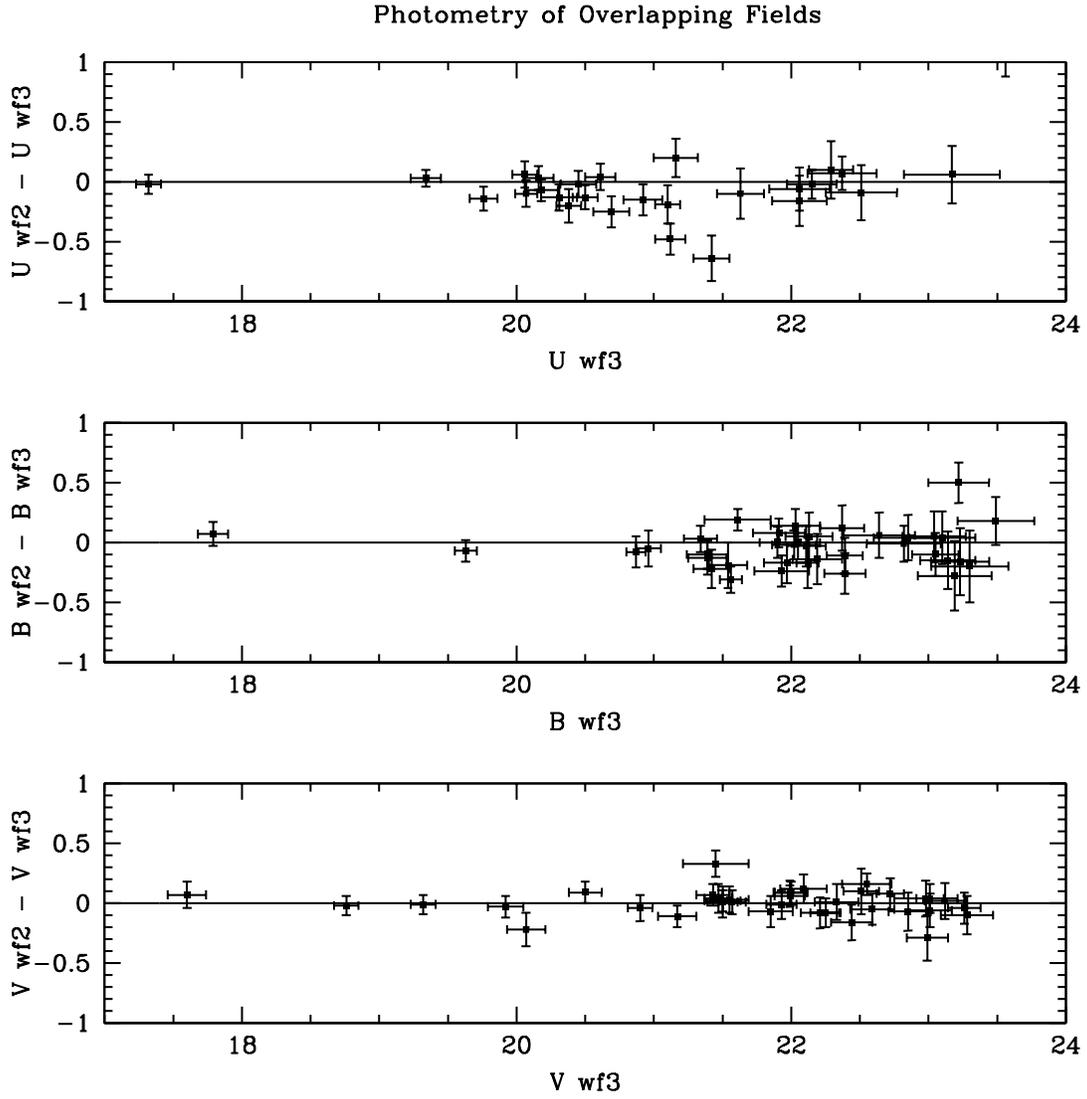,height=6.0in,angle=0}}
\caption{Residuals of photometry performed on the same stars in different fields in the F336W (U), F439W (B), and F555W (V), filters.  The residuals of the independent measurements are consistent with zero.  No systematic offest is seen.} 
\end{figure}

\begin{figure} 
\centerline{\psfig{file=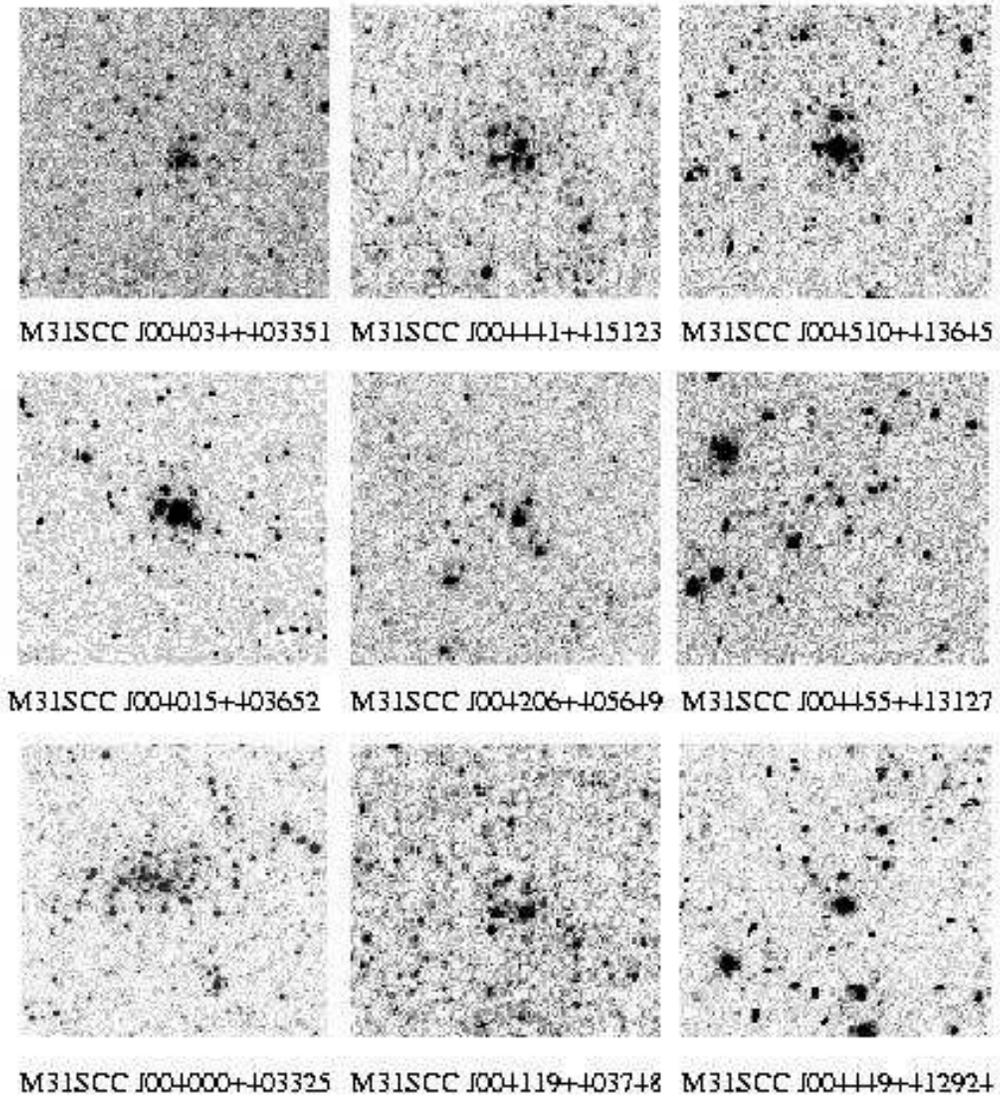,height=6.0in,angle=0}}
\caption{A random selection of 9 of our open cluster candidates.  These B band (F439W) images are 12'' by 12''.} 
\end{figure}

\begin{figure} 
\centerline{\psfig{file=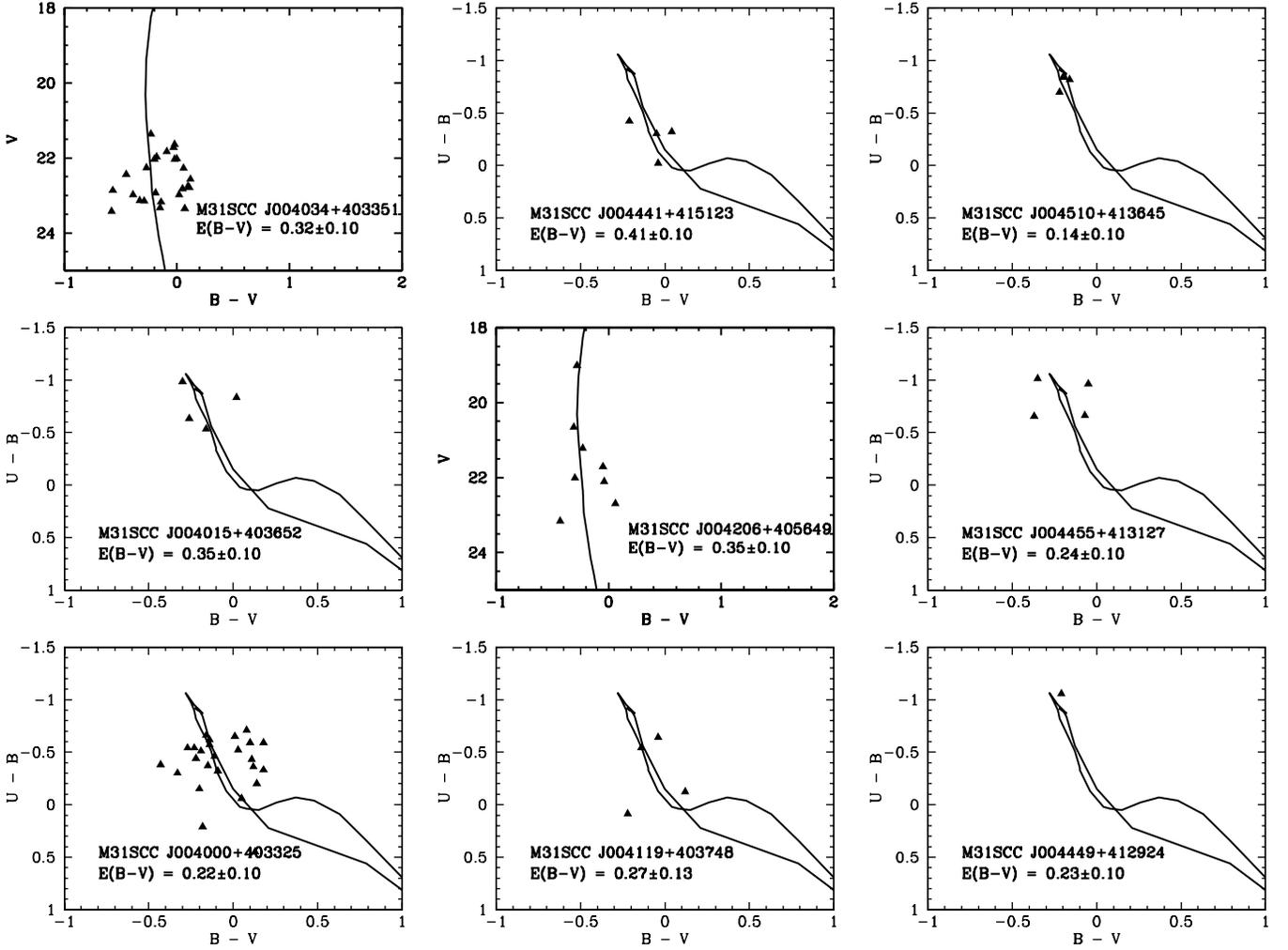,height=6.0in,angle=270}}
\caption{Reddening estimates of the 9 example clusters.  The stellar colors and magnitudes have been corrected by the reddening values shown on the figures.  These values gave the best fits to model upper main sequence colors.  In cases where we did not have U band photometry, we relied on the B-V colors alone to determine reddening.  Due to the sensitivity of WFPC2 in the U band, we occasionally were forced to make our first estimate of the reddening based on the colors of a single star detected in all three bands.} 
\end{figure}

\begin{figure} 
\centerline{\psfig{file=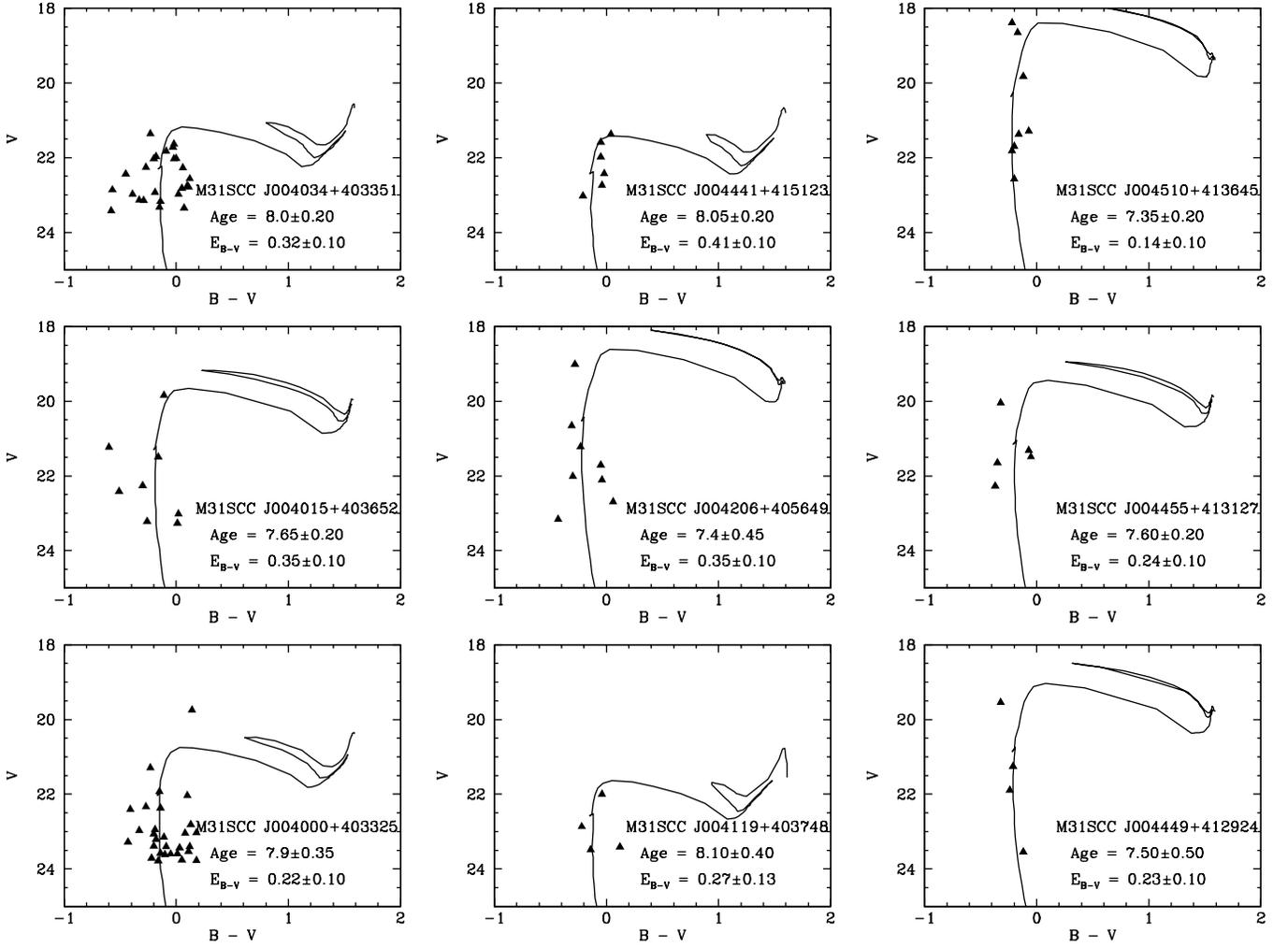,height=6.0in,angle=270}}
\caption{Age determinations of the 9 example clusters.  The ages were determined by eye to fit the turnoff to the bright blue stars in the cluster.  With so few stars, these ages estimates are very rough.} 
\end{figure}

\begin{figure} 
\centerline{\psfig{file=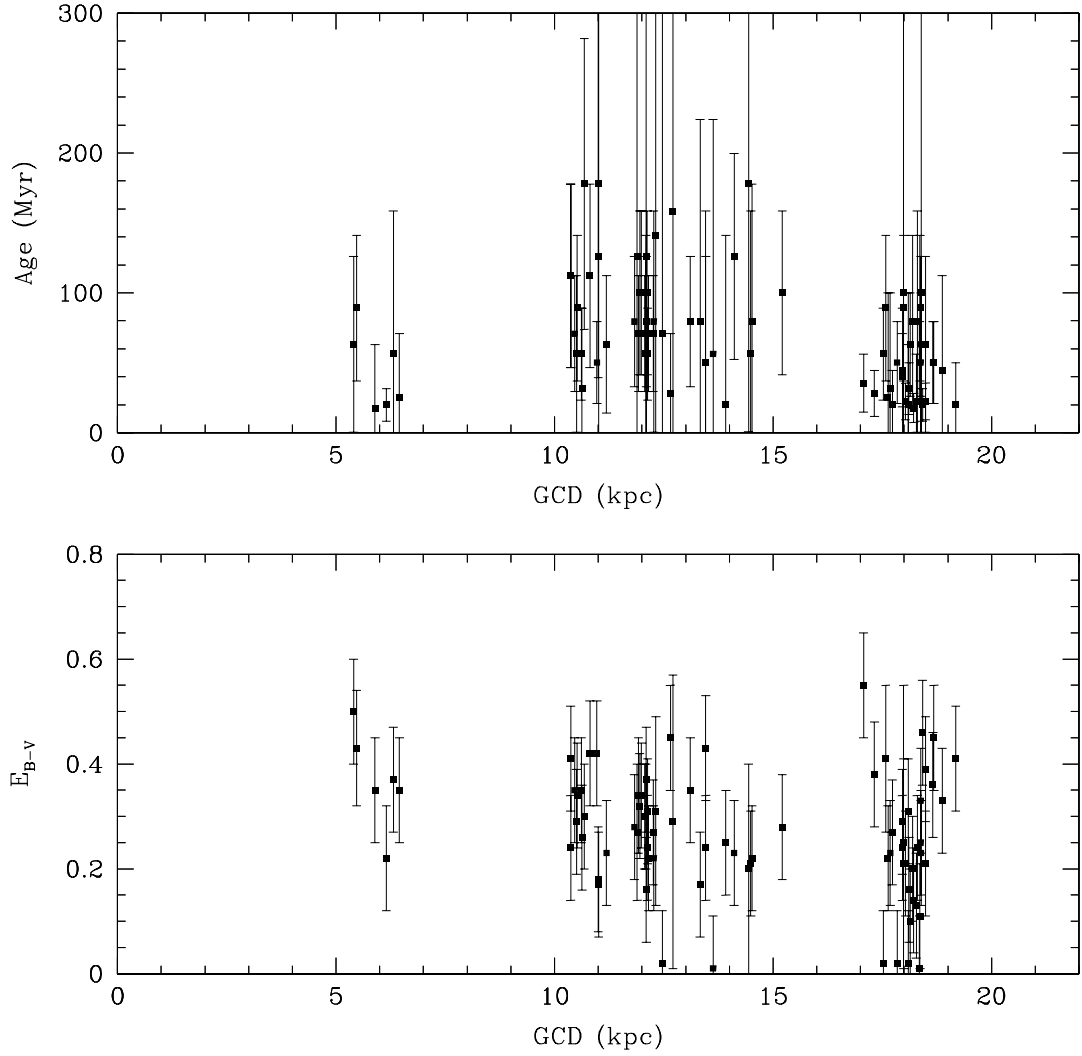,height=6.0in,angle=0}}
\caption{Plots of reddening and age vs. galactocentric distances for the cluster candidates.  No patterns are seen with these rough estimates.} 
\end{figure}

\begin{figure} 
\centerline{\psfig{file=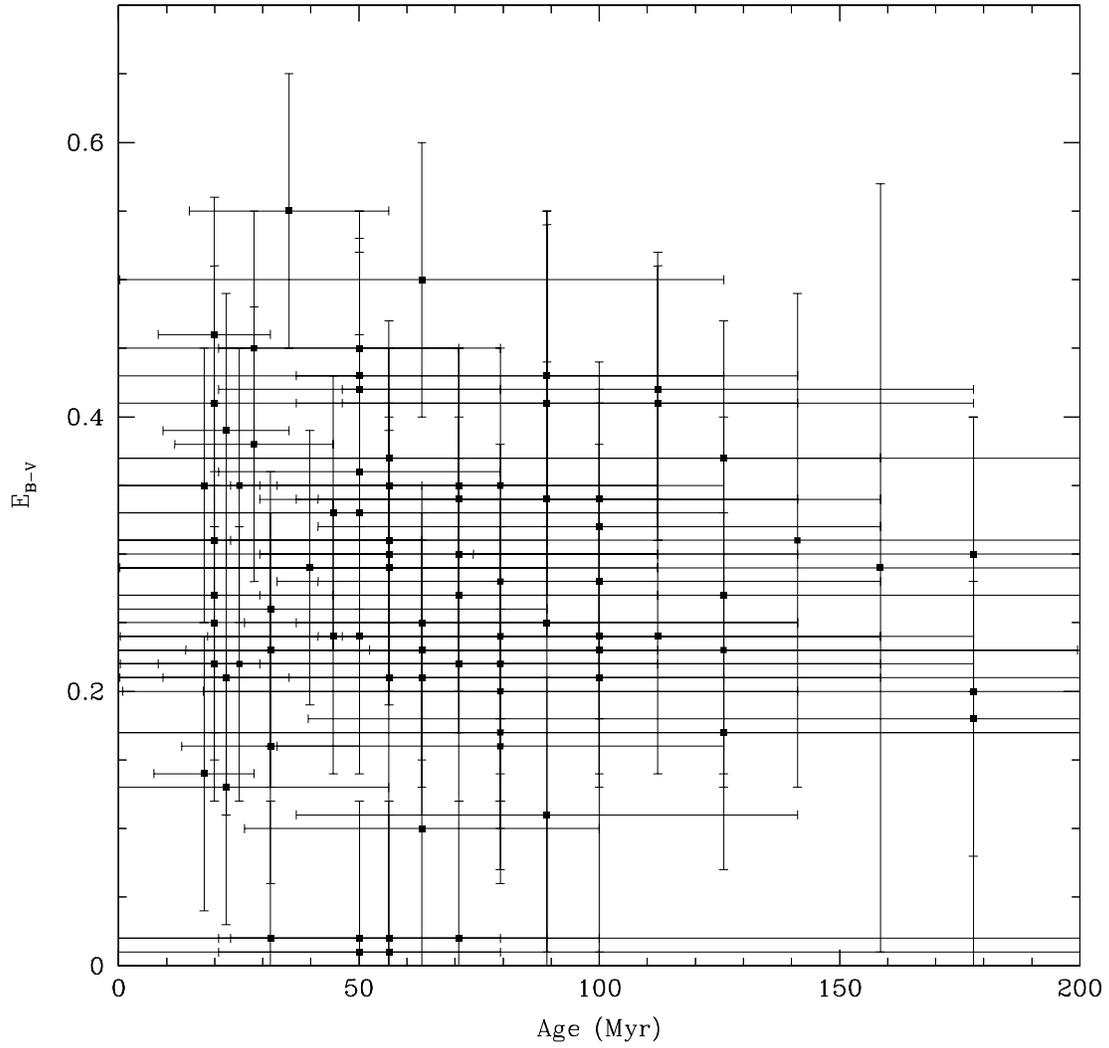,height=6.0in,angle=0}}
\caption{Reddening vs. age for the cluster candidates.  Our algorithm was not sensitive to red clusters of stars so that there is a lack of high extinction, old clusters.  The very low extinction clusters are likely to have field contaminants.} 
\end{figure}

\end{document}